\documentclass[12pt]{iopart}

\usepackage{graphicx}
\usepackage{hyperref}

\begin{document}

\title[Relativistic coupling of internal and center of mass dynamics]{Relativistic coupling of internal and center of mass dynamics in classical and simple bound quantum mechanical  systems}

\author{Dennis E. Krause}
\address{Department of Physics, Wabash College, Crawfordsville, IN  47933, USA}
\address{Department of Physics and Astronomy, Purdue University, West Lafayette, IN  47907, USA}

\author{Inbum Lee}
\address{Physics Department, Indiana University, Bloomington, Indiana 47405, USA}

\ead{kraused@wabash.edu}
\vspace{10pt}
\begin{indented}
\item[]\today
\end{indented}

\begin{abstract}
Although special relativity and quantum mechanics revolutionized physics in the early 20th century, the consequences of combining these two theories are still being explored a hundred years later, usually using the formidable theoretical machinery of quantum field theory. However,  a formalism accessible to undergraduates has been recently developed which shows how the center of mass and internal dynamics of classical and quantum systems is relativistically coupled with interesting  consequences. Here we  explore some of the implications of this coupling, first classically, where we find that the dynamics of the system is time dilated when moving relative to another inertial frame.  We then apply the  dynamics to a quantum 2-level atom bound in a 1-dimensional infinite potential well, and show that the coupling produces collapses and revivals in quantum interference.  This example provides an illustration of how the combination of special relativity and quantum mechanics can be studied in situations familiar to most undergraduates.
\end{abstract}

%
\vspace{2pc}

\vspace{2pc}
\noindent{Journal Reference: Eur.\ J.\ Phys.\ {\bf 38} (2017) 045401 \\ \url{https://doi.org/10.1088/1361-6404/aa6903}}
%
\maketitle
%
%

\section{Introduction}

Einstein's discovery that a body's rest energy $E_{\rm rest}$ is related to its inertial mass $m$ by
\begin{equation}
E_{\rm rest} = mc^{2}
\label{E = mc2}
\end{equation}
where $c$ is the speed of light, is arguably among the most famous in physics.  While most treatments of relativistic mechanics deal with particles,  Einstein is clear in his original paper that Eq.~(\ref{E = mc2}) is meant to be applied to any body, and speculates that the formula can be tested by searching for mass changes in materials with variable internal energy content such as radium salts \cite{Einstein 1905}.   As a consequence,  for a system with internal dynamics, one should expect that the internal motions would affect the center of mass motion through its effect on the system's inertial mass.  A container of gas at room temperature has a larger mass than the same gas at a lower temperature since the warmer gas has greater internal energy from the increased average kinetic energy of the gas molecules.  Yet, one does not  find much discussion in textbooks of how Eq.~(\ref{E = mc2}) might modify the dynamics of a system as compared to a particle. Presumably this is because the effects are tiny in most practical situations and the energy is a constant of motion. 

When a bright undergraduate attempts to raise similar questions in the quantum realm, an interesting problem arises.  In the usual time-dependent Schr\"{o}dinger wave equation for the wave function $\Psi(\vec{r},t)$ of a free particle encountered in quantum mechanics courses,
\begin{equation}
i\hbar\frac{\partial\Psi(\vec{r},t)}{\partial t} = -\frac{\hbar^{2}}{2m}\nabla^{2}\Psi(\vec{r},t),
\label{S eqn}
\end{equation}
the particle's mass $m$ is a parameter.  If one considers the system to be an atom, then the center of mass motion of the atom would be described by Eq.~(\ref{S eqn}), but now what should we use for the mass?  If we take Einstein seriously, then the mass should include the internal energy of the atom, but a quantum atom might not have a definite energy if it is in a superposition of internal energy states.  If the universe obeyed Galilean relativity, then Bargmann's theorem forbids a quantum system from existing in a superposition of mass states \cite{Bargmann,Kaempffer,Greenberger AJP,Greenberger PRL}, which would be inconsistent with Einstein.  Of course, as Einstein showed, our universe does not obey Galilean relativity, but  obeys special relativity (if we exclude gravity as we do here).  Even in the non-relativistic limit, features of the non-Galilean nature of the universe remain and void Bargmann's argument \cite{Greenberger PRL}.Yet, there still remains the problem of how to describe quantum mechanically the motion of the center of mass of a system in a superposition of internal energy states.

One might think that quantum field theory would be needed to address this problem, but recently several authors have developed a natural approach for incorporating the internal dynamics of a system into the dynamics of the system's center of mass by essentially making the mass of the system a dynamical variable \cite{Zych 2011,Zych 2012,Zych 2015,Pikovski Nature Physics,Pikovski Q&A,Zych 2016,Pikovski Time Dilation,Zych PhD}.   They have shown that in the quantum realm, this idea leads to special and general relativistic entanglement between a system's internal and external dynamics, producing  a loss of coherence in quantum twin paradox interference experiments \cite{Zych 2011} and universal decoherence from gravitational time dilation \cite{Pikovski Nature Physics}. (However, see Ref.~\cite{Bonder} for a criticism of the latter effect and Ref.~\cite{Pang,Toros} for  alternative interpretations.) The general principle at work is that a quantum system with internal dynamics (a ``quantum clock'') can be produced in a superposition of states that experience different proper times, leading to novel interference effects.

 In this paper, we will examine this work with a simple derivation of the Hamiltonian motivated by Einstein's mass-energy relation Eq.~(\ref{E = mc2}) which incorporates the internal and center of mass motions of the systems while explaining the limitations of the formalism.  A more formal derivation using Lagrangian mechanics is relegated to the Appendix.  Before applying the formalism to quantum mechanical systems, we first test  it out  classically, applying Hamilton's equations to the new Hamiltonian and showing that they lead to time-dilated dynamics.  We then extend the formalism to the quantum realm where we apply it to a 2-level atom trapped in an infinite potential well, a system involving topics familiar to any undergraduate who has taken quantum mechanics.  This simple system illustrates the consequences of the relativistic entanglement of internal and center of mass dynamics at a level appropriate for an upper-level undergraduate quantum mechanics course.

\section{System Hamiltonian}

In this section, we provide a simple derivation of the formalism developed in Refs.~\cite{Zych 2011,Zych 2012,Zych 2015,Pikovski Nature Physics,Pikovski Q&A,Zych 2016,Pikovski Time Dilation, Zych PhD}  for a small isolated system of interacting particles that are non-relativistic relative to their center of mass frame.  In the Appendix a more formal derivation is given of the system Hamiltonian using Lagrangian mechanics.   

We begin with the relativistic Hamiltonian describing a free point particle of mass $m$ with momentum $\vec{P}$ \cite{Goldstein}:
\begin{equation}
H_{\rm particle} = \sqrt{m^{2}c^{4} + P^{2}c^{2}}.
\label{particle H}
\end{equation}
  Let us now replace the particle with a very small compact system such as an atom or molecule of mass $M$.  Relative to the system's center of mass, the total momentum of the system vanishes, so the total energy of the system  $E_{\rm rest} = Mc^{2}$ in this frame (the rest energy) is an invariant quantity:
\begin{equation}
E_{\rm rest}  \equiv \left(c\sum_{\mu,\nu = 0}^{3}\eta_{\mu\nu}P^{\mu}_{\rm tot}P^{\nu}_{\rm tot}\right)^{1/2} = \sqrt{E_{\rm tot}^{2} - P_{\rm tot}^{2}c^{2}}.
\end{equation}
Here $P^{\mu}_{\rm tot} = (E_{\rm tot}/c, \vec{P}_{\rm tot})$ is the total 4-momentum of the system, and  $\eta_{\mu\nu} = {\rm diag}(1,-1,-1,-1)$ is the flat spacetime metric tensor.  Thus, the total energy relative to an arbitrary inertial frame can be written as
\begin{equation}
E =  \sqrt{E_{\rm rest}^{2} + P^{2}c^{2}},
\label{E}
\end{equation}
where from now on we will write $\vec{P}_{\rm tot} = \vec{P}$.

We now make the {\em ansatz} that the total Hamiltonian of the system can be obtained from Eq.~(\ref{E}) by making the replacement
\begin{equation}
E_{\rm rest} = Mc^{2} \rightarrow M_{0}c^{2} + H_{\rm int},
\label{replacement}
\end{equation}
where  $M_{0}$ is a constant, and $H_{\rm int}$ is the system's Hamiltonian relative to the center of mass frame, which, by definition, is written in terms of only the internal coordinates.   Then  $M_{0}c^{2}$ represents the non-dynamical part of the rest energy (e.g., the sum of the rest energies of the constituent particles).  In order for the concept of center of mass to make sense here, we assume that there exists a frame in  which all of the system particles are non-relativistic so that a center of mass frame can be defined \cite{Pryce}.  (See Ref.~\cite{Toros} for more discussion of the issue of center of mass in this context.)  This requires that we assume
\begin{equation}
\frac{H_{\rm int}}{M_{0}c^{2}} \ll 1,
\end{equation}
 for all subsequent calculations, and that we retain only first order terms in $H_{\rm int}/M_{0}c^{2}$.  With these assumptions, the total Hamiltonian of the freely propagating system becomes
\begin{equation}
H_{0}(\vec{P},q_{j},p_{j}) = \sqrt{M_{0}^{2}c^{4} + P^{2}c^{2}} + \frac{M_{0}c^{2}H_{\rm int}(q_{j},p_{j})}{\sqrt{M_{0}^{2}c^{4} + P^{2}c^{2}}},
\label{H0}
\end{equation}
where $\vec{P}$ is the total momentum, and $q_{j}$ and $p_{j}$ are the generalized coordinates and momenta relative to the center of mass.

If the system is under the influence of a time-independent external potential $U_{\rm ext}(\vec{r})$ which is sufficiently uniform over the size of the system $\ell$ that
\begin{equation}
\frac{\ell \, |\vec{\nabla}U_{\rm ext}(\vec{R})|}{|U_{\rm ext}(\vec{R})|} \ll 1,
\end{equation}
where $\vec{R}$ is the position of the center of mass, the effect of $U_{\rm ext}(\vec{r})$ should  depend only on $\vec{R}$ and be independent of the internal state of the system.  Assuming this condition, we write  the Hamiltonian of the system in the presence of this external interaction as
\begin{eqnarray}
H(\vec{R}, \vec{P},q_{j},p_{j}) &=&  H_{0} + U_{\rm ext}(\vec{R}) \nonumber \\
& = & \sqrt{M_{0}^{2}c^{4} + P^{2}c^{2}} + U_{\rm ext}(\vec{R}) + \frac{M_{0}c^{2}H_{\rm int}(q_{j},p_{j})}{\sqrt{M_{0}^{2}c^{4} + P^{2}c^{2}}}. 
\label{H}
\end{eqnarray}
The first two terms on the right side of Eq.~(\ref{H}) represent the special relativistic Hamiltonian of a point particle of mass $M_{0}$ and momentum $P$ in an external potential $U_{\rm ext}(\vec{R})$ \cite{Goldstein}.
 The third term on the right side of Eq.~(\ref{H}) contains both center of mass and internal coordinates, and represents the relativistic coupling  of the center of mass and internal dynamics.  Note that we have taken a conventional approach to incorporating a potential into the problem, which explicitly breaks Lorentz invariance without affecting the mass of the system.  One can find alternative approaches, where the potential produces a position-dependent mass (e.g., Ref.~\cite{Harvey}), but these go beyond the scope of this paper.  In addition, if the external potential varies significantly over the size of the system, additional coupling terms will arise \cite{Symon}, but these non-relativistic couplings are not of interest here.

 For the rest of the paper, we will explore the consequences of Eq.~(\ref{H}) that arise from the third term which couples the center of mass motion with its internal dynamics.

\section{Classical Dynamics}

We begin our investigation of the Hamiltonian given by  Eq.~(\ref{H}) by first applying classically Hamilton's equations to the center of mass momentum and position coordinates.  The momentum equation for the $i$th component gives
\begin{equation}
\frac{dP_{i}}{dt} =-\frac{\partial H}{\partial X_{i}} =  -\frac{\partial U_{\rm ext}}{\partial X_{i}}, 
\end{equation}
or 
\begin{equation}
\frac{d\vec{P}}{dt} = -\vec{\nabla}U_{\rm ext}(\vec{R}),
\end{equation}
which is the usual result from Newton's 2nd law when an  external force acts on the system.  Similarly, the equation for the system's center of mass velocity component $V_{i} = \dot{X}_{i} \equiv dX_{i}/dt$, where $X_{i}$ is the $i$th component of the system's center of mass position, is
\begin{equation}
\frac{dX_{i}}{dt} = \frac{\partial H}{\partial P_{i}},
\end{equation}
which gives for the expression for the center of mass velocity
\begin{equation}
\vec{V} = \frac{\vec{P}c^{2}}{\sqrt{M_{0}^{2}c^{4} + P^{2}c^{2}}}\left(1-  \frac{M_{0}c^{2}H_{\rm int}}{M_{0}^{2}c^{4} + P^{2}c^{2}}\right). 
\label{V}
\end{equation}
Eq.~(\ref{V}) is exactly what one would expect from the usual point particle formula,
\begin{equation}
\vec{V}_{\rm point} = \frac{\vec{P}c^{2}}{E} = \frac{\vec{P}c^{2}}{\sqrt{M^{2}c^{4} + P^{2}c^{2}}},
\end{equation}
after making the replacement Eq.~(\ref{replacement}) and retaining only the leading order term in $H_{\rm int}/M_{0}c^{2}$.

  Applying Hamilton's equations to the internal coordinates, we find
\begin{eqnarray}
\frac{dq_{i}}{dt} & = & \frac{\partial H}{\partial p_{i}} = \frac{M_{0}c^{2}}{\sqrt{M_{0}^{2}c^{4} + P^{2}c^{2}}} \frac{\partial H_{\rm int}}{\partial p_{i}}, \label{q dot}\\
\frac{dp_{i}}{dt} & = & -\frac{\partial H}{\partial q_{i}} = -\frac{M_{0}c^{2}}{\sqrt{M_{0}^{2}c^{4} + P^{2}c^{2}}} \frac{\partial H_{\rm int}}{\partial q_{i}} \label{p dot}.
\end{eqnarray}
Since
\begin{equation}
\sqrt{1 - V^{2}/c^{2}} =  \frac{M_{0}c^{2}}{\sqrt{M_{0}^{2}c^{4} + P^{2}c^{2}}} + {\cal O}\left(\frac{H_{\rm int}}{M_{0}c^{2}}\right),
\end{equation}
Eqs.~(\ref{q dot}) and (\ref{p dot}) are the  time-dilated Hamilton's equations viewed in the co-moving frame:
\begin{eqnarray}
\frac{dq_{i}}{d\tau} & = & \frac{\partial H_{\rm int}}{\partial p_{i}}, \\
\label{dq}
\frac{dp_{i}}{d\tau} & = & -\frac{\partial H_{\rm int}}{\partial q_{i}},
\label{dp}
\end{eqnarray}
where $d\tau$ is the proper time interval
\begin{equation}
d\tau = dt\sqrt{1 - V^{2}/c^{2}}.
\end{equation}
Therefore, the Hamiltonian given by Eq.~(\ref{H}) leads to physically intuitive and reasonable results for classical systems.   It also provides a more general underpinning for dynamical approaches of obtaining time dilation through the analysis of the physical mechanisms of clocks as done by Jefimenko \cite{Jefimenko}. Let us now turn our attention  to the dynamics of a quantum system governed by Eq.~(\ref{H}).

\section{Quantum Dynamics}

In this section, we will quantize the Hamiltonian Eq.~(\ref{H}).  Previous authors have obtained this Hamiltonian and used it to  examine the quantum interference of essentially freely propagating systems in gravitational fields \cite{Zych 2011,Zych 2012,Zych 2015,Pikovski Nature Physics,Pikovski Q&A,Zych 2016,Pikovski Time Dilation,Zych PhD}.  Here we will show that similar quantum interference arises in the simplest bound system encountered by undergraduate students: a 2-level atom in a 1-dimensional infinite potential well.  Not only is this system simple, it also can be solved exactly.  For novel behavior to be observed, the system will then be prepared in a superposition of both internal and center of mass states.  Using the density operator formalism to trace over the internal states, the center of mass position probability density is obtained, revealing a collapse and revival of the center of mass position  interference due to the relativistic coupling.

\subsection{Hamiltonian Operator}

In our analysis of the classical system, we assumed that  one can find a frame in which the system is non-relativistic with a well-defined center of mass.  We then placed essentially no restrictions on the motion of this center of mass.  In this paper, we wish to investigate the quantum aspects of our problem without having to deal with the significant complications of particle creation, which  requires a treatment using quantum field theory.  Therefore, we will now assume $P^{2}/M_{0}^{2}c^{2} \ll 1$ and include only the leading order special relativistic terms of our Hamiltonian Eq.~(\ref{H}), which now becomes
\begin{eqnarray}
H(\vec{R}, \vec{P},q_{j},p_{j}) &\simeq & M_{0}c^{2} + \frac{P^{2}}{2M_{0}} -  \frac{P^{4}}{8M_{0}^{3}c^{2}} + U_{\rm ext}(\vec{R}) \nonumber \\
&& \mbox{} + H_{\rm int}(q_{j},p_{j})\left(1 - \frac{P^{2}}{2M_{0}^{2}c^{2}}\right).
\label{NR H}
\end{eqnarray}

So far our treatment has been classical and three-dimensional.  For the rest of this paper, we will focus our attention on 1-dimensional quantum mechanical systems, where we replace the center of mass canonical coordinates $X$ and $P_{x}$ with their corresponding quantum mechanical operators $\hat{X}$ and $\hat{P}_{x}$, while making similar substitutions with the internal coordinates.  This leads to the quantum mechanical Hamiltonian describing a small quantum system bound to a potential,
\begin{eqnarray}
\hat{H} & \simeq &  M_{0}c^{2} + \frac{\hat{P}_{x}^{2}}{2M_{0}} -  \frac{\hat{P}_{x}^{4}}{8M_{0}^{3}c^{2}} + U_{\rm ext}(\hat{X})  \nonumber \\
&& \mbox{} + \hat{H}_{\rm int}(\hat{q}_{j},\hat{p}_{j})\left(1 - \frac{\hat{P}_{x}^{2}}{2M_{0}^{2}c^{2}}\right).
\label{NR X H}
\end{eqnarray}
To account for ground state energies, the constant $M_{0}$ is defined so that the energy of the internal system vanishes when it is in its ground state.

We note in passing that all of these results could have been obtained by starting with the single particle Hamiltonian
\begin{equation}
\hat{H} = Mc^{2} + \frac{\hat{P_{x}}^{2}}{2M} -  \frac{\hat{P}_{x}^{4}}{8M^{3}c^{2}} + U_{\rm ext}(\hat{X}),
\label{normal H}
\end{equation}
and replacing the particle mass $M$ with the mass operator $\hat{M}$ defined by \cite{Zych 2015}
\begin{equation}
\hat{M} \equiv M_{0} + \frac{\hat{H}_{\rm int}}{c^{2}},
\end{equation}
and keeping only first order terms in $\hat{H}_{\rm int}/M_{0}c^{2}$.  In essence, this takes Einstein's famous formula Eq.~(\ref{E = mc2}) seriously in the quantum realm where systems can be in superpositions of energy states, which also means they should be in superpositions of mass eigenstates.   Arguments have been made against mass superpositions \cite{Bargmann,Kaempffer,Greenberger AJP,Greenberger PRL},  but they are based on Galilean relativity rather than the non-relativistic limit of special relativity \cite{Pikovski Q&A}.  Finally, one can similarly deal with unstable quantum particles by replacing $M$ in Eq.~(\ref{normal H}) with a complex mass \cite{Krause PLA,Krause arXiv}
\begin{equation}
\tilde{M} \equiv M_{0} - \frac{i\hbar\Gamma_{0}}{2},
\end{equation}
where $\Gamma_{0}$ is the decay rate of the particle in its rest frame,  and keeping only leading order terms in $\hbar\Gamma_{0}/M_{0}c^{2}$.  In both cases, the system with an internal Hamiltonian $\hat{H}_{\rm int}$ and the unstable particle with width $\Gamma_{0}$, the quantum system has indefinite mass.

In practice, one frequently considers problems where the non-relativistic center of mass and internal dynamics are known, and the remaining portions are small in comparison. To use perturbation theory in this case,  it is helpful to separate the Hamiltonian Eq.~(\ref{NR X H}) into two parts,
\begin{equation}
\hat{H} = \hat{H}_{0} + \hat{H}_{1}.
\end{equation}
Here the unperturbed Hamiltonian,
\begin{equation}
\hat{H}_{0} =  \frac{\hat{P}_{x}^{2}}{2M_{0}} + U_{\rm ext}(\hat{X}) + M_{0}c^{2} + \hat{H}_{\rm int},
\end{equation}
 can also be written as the sum of two parts:
 \begin{equation}
 \hat{H}_{0}  = \hat{H}_{0,{\rm cm}}  + \hat{H}_{0,{\rm int}}, 
 \end{equation}
 where the center of mass portion,
 \begin{equation}
 \hat{H}_{0,{\rm cm}} =  \frac{\hat{P}_{x}^{2}}{2M_{0}} + U_{\rm ext}(\hat{X})
 \label{H0 cm}
 \end{equation}
 is the usual non-relativistic quantum Hamiltonian for a particle in an external potential, and
 \begin{equation}
 \hat{H}_{0,{\rm int}} =   M_{0}c^{2} + \hat{H}_{\rm int},
 \end{equation}
 is the usual internal Hamiltonian relative to the center of mass (including rest energies of the component particles). The small perturbation is given by
\begin{equation}
\hat{H}_{1} = -\frac{\hat{P}_{x}^{4}}{8M_{0}^{3}c^{2}} -  \frac{\hat{P}_{x}^{2}\hat{H}_{\rm int}}{2M_{0}^{2}c^{2}},
\label{H1}
\end{equation}
where the first term leads to small corrections to the center of mass motion, while the second term will couple the center of mass and internal dynamics.

\subsection{System: 2-Level Atom in Infinite Potential Well}

So far our system has  been left unspecified.  Let us now focus our attention on the simplest possible system that will exhibit the desired features: a 2-level atom.  The eigenvalue equation for $\hat{H}_{0,{\rm int}}$ will be 
\begin{equation}
\hat{H}_{0,{\rm int}}|n\rangle = E_{n}^{\rm int}|n\rangle,
\label{eigen 2-level}
\end{equation}
where $n = 0, 1$,  $E_{0}^{\rm int} = 0$ is the ground state energy, and $E_{1}^{\rm int}$ is the energy of the excited state.  

The simplest bound system is a particle in a 1-dimensional infinite potential well of length $L$ given by
\begin{equation}
U_{\rm ext}(X)= U_{\rm well}(X) = \left\{\begin{array}{ll}
 0, & \mbox{if $0 < X < L$}, \\
\infty, &  \mbox{otherwise}.
\end{array}\right.
\end{equation}
In this case, the eigenvalue equation of the unperturbed center of mass motion Hamiltonian Eq.~(\ref{H0 cm}) is \cite{Griffiths IPW}
\begin{equation}
\hat{H}_{0,{\rm cm}}|N\rangle = E_{N}^{\rm IP}|N\rangle,
\label{eigen infinite well}
\end{equation}
where 
\begin{equation}
E_{N}^{\rm IP} = \frac{(N+1)^2 \pi^{2} \hbar^{2}}{2M_{0}L^{2}}, \,\,\,N = 0, 1, 2, \ldots 
\end{equation}
and the energy wave functions are
\begin{equation}
\psi_{N}^{\rm IP}(X) = \langle X|N\rangle = \left\{\begin{array}{ll}
\displaystyle \sqrt{\frac{2}{L}}\sin\left[\frac{(N+1)\pi x}{L}\right], & \mbox{if $0 < X < L$}, \\
& \\
0 &  \mbox{otherwise}.
\end{array}\right.
\label{IP wave functions}
\end{equation}
(We've chosen the unconventional notation of numbering the ground state $N = 0$ instead of $N = 1$ so that both the center of mass and internal Hamiltonian ground states in this paper will be labeled with $N = n = 0$.)  Combining the results of Eqs.~(\ref{eigen 2-level}) and (\ref{eigen infinite well}), the total unperturbed eigenvalue equation is
\begin{equation}
\hat{H}_{0}|N,n\rangle = E^{0}_{Nn}|N,n\rangle,
\label{eigen total infinite well}
\end{equation}
where the  energy of the unperturbed system is 
\begin{equation}
E_{Nn}^0= E_{N}^{\rm IP} + E_{n}^{\rm int}  = \frac{(N+ 1)^2 \pi^2 \hbar^2}{2m_0L^2}+E_{n}^{\rm int}.
\end{equation}

We are fortunate that for this system that the unperturbed Hamiltonian and the perturbation commute, $[\hat{H}_{0},\hat{H}_{1}] = 0$, so that the eigenstates of $\hat{H}_{0}$ are also eigenstates of  $\hat{H}_{1}$, and thus, of the Hamiltonian $\hat{H}$,:
\begin{equation}
|E_{Nn}\rangle = |N,n\rangle.
\end{equation}
  The energy contribution from $\hat{H}_{1}$ is then
\begin{eqnarray}
E_{Nn}^1 & = & \langle N,n|\hat{H}_{1} |N,n\rangle \nonumber \\
	& = &-\langle N,n|\left(\frac{\hat{P}_{x}^{4}}{8M_{0}^{3}c^{2}} +  \frac{\hat{P}_{x}^{2}\hat{H}_{\rm int}}{2M_{0}^{2}c^{2}}\right)|N,n\rangle \nonumber \\
	&=&-\left[\frac{1}{8M_{0}^{3}c^{2}}\langle N |\hat{P}_{x}^{4}|N\rangle
		+ \frac{E_{n}^{\rm int}}{2M_{0}^{2}c^{2}} \langle N |\hat{P}_{x}^{2}|N\rangle \right]. \nonumber \\
\label{E1Nn 1}
\end{eqnarray}
Using the infinite potential well wave functions given by Eq.~(\ref{IP wave functions}),
\begin{eqnarray}
\langle N |\hat{P}_{x}^{2}|N\rangle & = & -\hbar^{2}\int^{L}_{0}dX\,\psi_{N}^{{\rm IP}*}(X)\frac{d^2\psi_{N}^{\rm IP}(X)}{dX^2}
	\nonumber \\
	& = & -\hbar^{2}\left[-\frac{\pi^{2}(N + 1)^{2}}{L^{2}}\right] \nonumber \\
	&& \mbox{}\times \frac{2}{L} ~\int^{L}_{0}dX\,\sin^2\left[\frac{(N+ 1) \pi X}{L} \right] \nonumber \\
	& = & \frac{(N + 1)^{2}\pi^{2}\hbar^{2}}{L^{2}} = 2M_{0}E_{N}^{\rm IP},
\end{eqnarray}
and
\begin{eqnarray}
\langle N |\hat{P}_{x}^{4}|N\rangle & = & \hbar^{4}\int^{L}_{0}dX\,\psi_{N}^{{\rm IP}*}(X)\frac{d^4\psi_{N}^{\rm IP}(X)}{dX^4}
	\nonumber \\
		& = & \frac{(N + 1)^{4}\pi^{4}\hbar^{4}}{L^{4}} = 4M_{0}^{2}(E_{N}^{\rm IP})^{2}.
\end{eqnarray}
Substituting these results into Eq.~(\ref{E1Nn 1}) gives
\begin{equation}
E^{1}_{Nn} = -\frac{(E_{N}^{\rm IP})^{2}}{2M_{0}c^{2}} - \frac{E_{N}^{\rm IP}E_{n}^{\rm int}}{M_{0}c^{2}},
\end{equation}
so the total energy of the system is
\begin{equation}
E_{Nn} = E_{N}^{\rm IP} + E_{n}^{\rm int}  - \frac{(E_{N}^{\rm IP})^{2}}{2M_{0}c^{2}} - \frac{E_{N}^{\rm IP}E_{n}^{\rm int}}{M_{0}c^{2}}.
\label{ENn}
\end{equation}
For a particle in an infinite potential well, $E^{\rm IP}_{N}$ is just the kinetic energy, so the first two terms on the right side of Eq.~(\ref{ENn}) are the usual non-relativistic kinetic and internal energies, while the third term is the leading order relativistic correction to the kinetic energy.  The fourth term on the right side of Eq.~(\ref{ENn}) is  the contribution to the kinetic energy due to the system's internal energy.

We now have everything needed to investigate the dynamics of this system.

\subsection{Dynamics}

The quintessential quantum phenomenon is interference, which arises when a quantum system exists in a coherent superposition of states, which is analogous to the coherent superposition of wave amplitudes that gives rise to classical wave interference \cite{Feynman,Lovett}.  Coherence is preserved as long is there is no way of determining which state the system is in.  If information about which state the system is in can be transferred to another system (i.e., the states of the two systems become entangled), coherence is lost and interference disappears.  

 In order to observe interference in the center of mass motion of our atom in the infinite potential well, the system needs to be placed into a superposition of center of mass energy states.  Furthermore, to investigate the effects of the system having an indefinite mass, the atom should also be in a superposition of internal energy states.  Therefore, we will assume the system begins at $t =0$ in state that is a tensor product of superpositions of center of mass and internal states,
\begin{eqnarray}
|\Psi(0)\rangle & = &  \left[\frac{1}{\sqrt{2}}\left(|0\rangle + |1\rangle\right)\right]_{\rm cm} \otimes \left[\frac{1}{\sqrt{2}}\left(|0\rangle + |1\rangle\right)\right]_{\rm int} \nonumber \\
	& = & \frac{1}{2}\left(|0,0\rangle + |0,1\rangle + |1,0\rangle + |1,1\rangle\right).
\end{eqnarray}
That is, this initial state is {\em not} entangled so knowing the internal state provides no information about the center of mass state, and vice versa.
Then for $t > 0$, the state vector for the system can be written down by including exponential factors $e^{-iE_{Nn}t/\hbar}$ since  $|N,n\rangle$ is an eigenstate of the total Hamiltonian:
\begin{eqnarray}
|\Psi(t)\rangle & = &  \frac{1}{2}\sum_{N=0}^{1}\sum_{n =0}^{1}e^{-iE_{Nn}t/\hbar}|N,n\rangle
	\nonumber \\
	& = & \frac{1}{2}\left(e^{-iE_{00}t/\hbar}|0,0\rangle + e^{-iE_{01}t/\hbar} |0,1\rangle \right. \nonumber \\
	& & \left. \mbox{} + e^{-iE_{10}t/\hbar}|1,0\rangle + e^{-iE_{11}t/\hbar} |1,1\rangle\right).
\end{eqnarray}
The system's density operator is then
\begin{equation}
\hat{\rho}(t) =  |\Psi(t)\rangle \langle\Psi(t)| = \frac{1}{4}\sum_{N,N'=0}^{1}\sum_{n,n' =0}^{1}e^{-i(E_{Nn}- E_{N'n'})t/\hbar}|N,n\rangle\langle N',n'|.
\label{rho t}
\end{equation}

Let us now focus our attention on the center of mass motion of the system.  Tracing Eq.~(\ref{rho t}) over the internal energy states gives the reduced density operator of the center of mass motion, which has matrix elements
\begin{eqnarray}
\langle K| \hat{\rho}_{\rm cm}(t)|K'\rangle & = & \sum_{k = 0}^{1}\langle K,k|\hat{\rho}(t)|K',k\rangle, \nonumber \\
		& = & \frac{1}{4}\sum_{N,N'=0}^{1}\sum_{n,n' =0}^{1}\sum_{k = 0}^{1}e^{-i(E_{Nn}- E_{N'n'})t/\hbar} \nonumber \\
		&& \mbox{} \times
		\langle K,k|N,n\rangle\langle N',n'|K',k\rangle \nonumber \\
				& = & \frac{1}{4}\left[e^{-i(E_{K0}- E_{K'0})t/\hbar} + e^{-i(E_{K1}- E_{K'1})t/\hbar}\right]. \nonumber
		\\
\end{eqnarray}
The diagonal matrix elements of $\rho_{\rm cm}(t)$ are constant,
\begin{equation}
\langle 0| \hat{\rho}_{\rm cm}(t)|0\rangle = \langle 1| \hat{\rho}_{\rm cm}(t)|1\rangle = \frac{1}{2},
\end{equation}
so there is always a 50\% probability of the center mass energy being in either of the lowest two energy states.  The interference in the center of mass motion arises from the off-diagonal elements:
\begin{eqnarray}
\langle 0| \hat{\rho}_{\rm cm}(t)|1\rangle & = & \langle 1| \hat{\rho}_{\rm cm}(t)|0\rangle ^{*} \nonumber \\
 & = &  \frac{1}{4}\left[e^{-i(E_{00}- E_{10})t/\hbar} + e^{-i(E_{01}- E_{11})t/\hbar}\right] \nonumber \\
& = & \frac{1}{2}e^{i\Omega_{\rm cm}t}\cos(\Omega_{\rm ent} t).
\label{rho cm}
\end{eqnarray}
We see that two natural frequencies, $\Omega_{\rm cm}$ and $\Omega_{\rm ent}$, arise.  First, there is the  frequency of the center of mass motion $\Omega_{\rm cm}$ that depends primarily on the energy difference of the two center of mass energy states,
\begin{eqnarray}
\Omega_{\rm cm} & \equiv & \frac{|E_{11}+ E_{10} - E_{01} - E_{00}|}{2\hbar}
\nonumber \\
& = & \frac{E_{1}^{\rm IP}}{\hbar}\left(1 - \frac{E_{1}^{\rm IP}}{2M_{0}c^{2}}\right) -  \frac{E_{0}^{\rm IP}}{\hbar}\left(1 - \frac{E_{0}^{\rm IP}}{2M_{0}c^{2}}\right) \nonumber \\
&& \mbox{} - \frac{1}{2\hbar M_{0}c^{2}}\left(E_{1}^{\rm IP}-E^{\rm IP}_{0}\right)\left(E_{0}^{\rm int}+E^{\rm int}_{1}\right)  \nonumber \\
& = &  \frac{(E_{1}^{\rm IP} - E_{0}^{\rm IP})}{\hbar}\left[1 - \frac{E_{0}^{\rm int} + E_{1}^{\rm int}}{2M_{0}c^{2}}\right] - \frac{1}{2\hbar M_{0}c^{2}}\left[(E_{1}^{\rm IP})^{2} 
- (E_{0}^{\rm IP})^{2}\right].
\end{eqnarray}
This result has a nice intuitive interpretation.  This frequency would arise for a particle in a superposition of the first two  infinite square well energy states if it had a  mass
\begin{equation}
M = M_{0} + \frac{\langle \hat{H}_{\rm int}\rangle}{c^{2}},
\end{equation}
where 
\begin{equation}
\langle \hat{H}_{\rm int}\rangle= \frac{1}{2}\left(E_{0}^{\rm int} + E_{1}^{\rm int}\right)
\end{equation}
is the average internal energy.
The second natural  frequency $\Omega_{\rm ent}$ which appears in Eq.~(\ref{rho cm}) arises purely from the coupling term of the Hamiltonian and depends on the product of the differences in center of mass and internal energy levels:
\begin{eqnarray}
\Omega_{\rm ent} & \equiv &\frac{|E_{11} + E_{00} - E_{10} - E_{01}|}{2\hbar} \nonumber \\
& = &  \frac{1}{2\hbar M_{0}c^{2}}
\left(E_{1}^{\rm IP}  - E_{0}^{\rm IP})(E_{1}^{\rm int}  - E_{0}^{\rm int}\right). 
\end{eqnarray}
We will  see  that this frequency is associated with the collapse and revival of the interference due to the  entanglement of the center of mass and internal motions.  This effect will not occur if the center of mass and/or the internal systems are in energy eigenstates.

As a demonstration of the effects of the two frequencies, let us look at the center of mass probability density, which is given by
\begin{eqnarray}
{\cal P}_{\rm cm}(X,t) & = & {\rm Tr}\left[\hat{\rho}_{\rm cm}(t)|X\rangle\langle X|\right] \nonumber \\
	& = & \sum_{K,K' = 0}^{1}\langle K| \hat{\rho}_{\rm cm}(t)|K'\rangle\langle K'|X\rangle \langle X|K\rangle 
		\nonumber \\
	& = & \sum_{K,K' = 0}^{1}\langle K| \hat{\rho}_{\rm cm}(t)|K'\rangle \psi_{K'}^{{\rm IP}*}(X)\psi_{K}^{\rm IP}(X)
		\nonumber \\
	& = & \frac{1}{2}|\psi_{0}^{\rm IP}(X)|^{2} +  \frac{1}{2}|\psi_{1}^{\rm IP}(X)|^{2} \nonumber \\
	&& \mbox{}
		+ \frac{1}{2} \cos(\Omega_{\rm ent}t)
	\left[
	\psi_{0}^{{\rm IP}*}(X)\psi_{1}^{\rm IP}(X) e^{i\Omega_{\rm cm}t}  \right.
	\nonumber \\
	&& \left.\mbox{}
	+ \psi_{0}^{\rm IP}(X)\psi_{1}^{{\rm IP}*}(X) e^{-i\Omega_{\rm cm}t} 
	\right].
\end{eqnarray}
 Since the center of mass wave functions are real, this simplifies to
\begin{eqnarray}
{\cal P}_{\rm cm}(X,t)  & = & \frac{1}{2} \left[ |\psi_{0}^{\rm IP}(X)|^{2} + |\psi_{1}^{\rm IP}(X)|^{2} \right. \nonumber \\
&& \left. \mbox{} + 2\psi_{0}^{\rm IP}(X)\psi_{1}^{\rm IP}(X) \cos(\Omega_{\rm ent}t) \cos(\Omega_{\rm cm}t) \right].
\label{P cm}
\end{eqnarray}
Eq.~(\ref{P cm}) has a form analogous to the probability density seen in the usual double-slit experiment \cite{Feynman} where, in that case, the wave functions represent the amplitudes of passing through the slits, except the   interference term here depends on time instead of spatial position.  

\begin{figure}[t!]
\centering
\includegraphics[width=10cm]{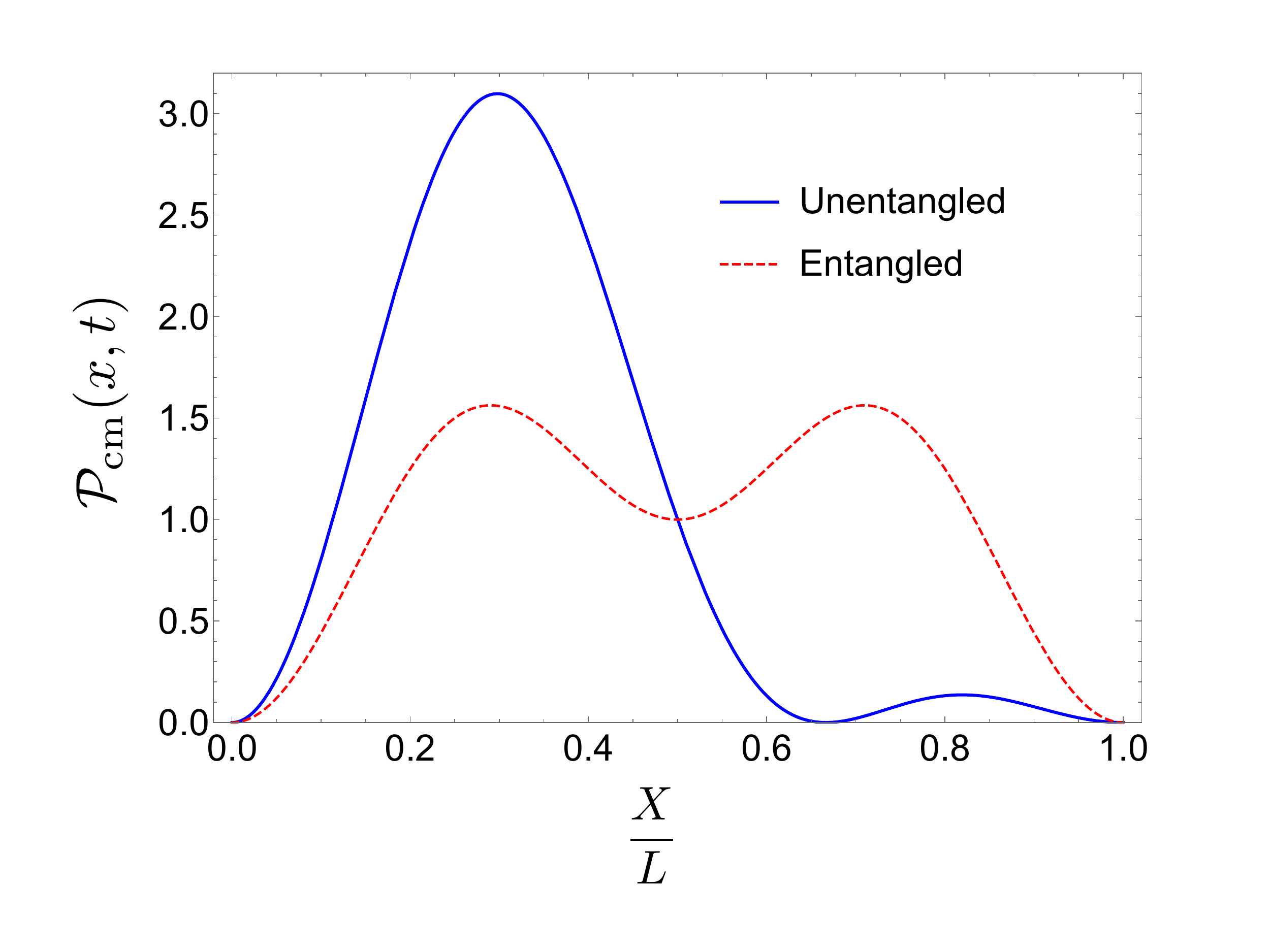}
\caption{Plots of  the center of mass probability density given by Eq.~(\ref{P cm}) at $t = 0$ (solid, blue line) when the center of mass and internal systems are completely unentangled (maximum interference), and at $t = \pi/2\Omega_{\rm net}$ (dashed, red line) when they are fully entangled (no interference).}
\label{Plot figure}
\end{figure}

At $t = 0$, $\cos(\Omega_{\rm ent}t) = 1$ so the center of mass probability density oscillates with frequency $\Omega_{\rm cm}$ in the same manner as a particle with no internal degrees of freedom.  As noted earlier, we've chosen to start the system completely unentangled so the center of mass is described by a pure state wave function, and the internal state contains no information about the center of mass motion.  However, as time passes,   $\cos(\Omega_{\rm ent}t)$ decreases as the center of mass and internal states become entangled through the relativistic coupling.  Conceptually, because all of the center of mass energy in the infinite square well is kinetic, the particle in the higher energy  $N =1$ center of mass  state is traveling faster than if was in the ground ($N = 0$) state, so the time dilation of the internal dynamics will differ.  By observing how much the internal clock has slowed, one could tell which center of mass state the  system is in.  This is how information about the center of mass state becomes encoded in the internal state of the atom.  Eventually, when $t = \pi/2\Omega_{\rm ent}$, the internal state has become completely entangled with the center of mass state, shutting off the interference.  The oscillation in the probability density ${\cal P}_{\rm cm}(X,t)$ slows to a stop.  At this instant, the information about the center of mass motion is completely encoded into the internal state.  The probability density is simply the equally weighted sum of the probability densities of being in the center of mass states $N = 0$ and $N = 1$, which is analogous to the double-slit interference when one knows through which slit the particle passes. Then, as time continues to advance, the system gradually loses its entanglement and interference is restored.  In Fig.~\ref{Plot figure}, we have plotted  ${\cal P}_{\rm cm}(X,t)$ for $t = 0$ (when the system is unentangled and exhibits maximum interference) and $t = \pi/2\Omega_{\rm ent}$ (when the system is fully entangled and there is no interference).
This oscillating interference due to $\cos(\Omega_{\rm ent}t)$ for the bound atom is analogous the oscillating visibility in quantum interferometry from an atom experiencing different general relativistic proper times while traversing an interferometer \cite{Zych 2011}.

\section{Discussion}

Most treatments of Einstein's formula Eq.~(\ref{E = mc2}) relating the inertial mass of a system to its rest energy consider the system to be a particle, while nearly all systems one  encounters in reality possess internal degrees of freedom. In quantum mechanics, systems may have indefinite internal energy if they are unstable or are in an internal state that is a superposition of energy eigenstates.    In textbook treatments, it is not obvious how to incorporate such an indefinite mass into the Schr\"{o}dinger wave equation even though Eq.~(\ref{E = mc2}) implies that it should be possible.

 In this paper, we explored the classical and quantum consequences of a new approach which takes Einstein seriously and incorporates the  internal degrees of freedom for systems which have a well-defined center of mass \cite{Zych 2011,Zych 2012,Zych 2015,Pikovski Nature Physics,Pikovski Q&A,Zych 2016,Pikovski Time Dilation,Zych PhD}. We presented both a simple and, in the Appendix, a more formal derivation of the Hamiltonian Eq.~(\ref{H}) which leads to the relativistic coupling of the internal and center of mass dynamics, and showed that for classical systems, this produces the expected time dilation of the dynamics of a system when viewed from another inertial frame. 
 
 When we quantized the formalism for nearly non-relativistic systems, we  found that the resulting Hamiltonian operator Eq.~(\ref{NR X H}) includes a term which couples the center of mass and internal dynamics.  This will lead to a change of dynamics and, using a 2-level atom bound in a 1-dimensional infinite potential well, we saw that it can produce an oscillation in the entanglement between the center of mass and internal motions, significantly altering the quantum interference compared to the situation of an analogous system with no internal degrees of freedom.

We hope that this paper will help address  questions that might arise from a curious student who wishes to go beyond typical textbook discussions of relativistic particles and thinks about systems with internal dynamics. The formalism developed recently \cite{Zych 2011,Zych 2012,Zych 2015,Pikovski Nature Physics,Pikovski Q&A,Zych 2016,Pikovski Time Dilation,Zych PhD} provides a natural approach consistent with Einstein's famous equation which leads to physically sensible classical dynamics and novel interference effects in the quantum realm.    It  provides an elegant way of studying the motion of a quantum system with indefinite mass due to being in an internal state that is a superposition of energy eigenstates. While we only considered applications involving special relativity in order to provide the simplest treatment, the formalism has been shown to apply more broadly to general relativistic systems. Finally, the formalism is  accessible to undergraduates, giving them an opportunity to explore fundamental issues in relativity and quantum mechanics that may soon be studied in the laboratory.

\appendix
\section{More Formal Derivation of System Hamiltonian}

In this appendix, we review a more complete derivation of the free system Hamiltonian, Eq.~(\ref{H}), that was outlined in Ref.~\cite{Pikovski Nature Physics}.  We begin with the action of a small system in a frame co-moving with its center of mass,
\begin{equation}
S = \int L_{\rm rest}(q_{i},q_{i}')\,d\tau,
\label{rest action}
\end{equation}
where $L_{\rm rest}(q_{i},q_{i}')$ is the system Lagrangian relative to this rest frame, $q_{i}$ and $q_{i}'$ are the generalized coordinates and velocities for the system relative to the center of mass frame.  We use primes to denote derivatives with respect to the system's proper time $\tau$, e.g., $q_{i}' = dq_{i}/d\tau$.    If the system is simply a point particle with mass $m$, $L_{\rm rest} = -mc^{2}$.

If we now view the system's motion relative to an inertial frame, then the action Eq.~(\ref{rest action}) can be written as 
\begin{equation}
S = \int L_{\rm rest}\left(q_{i},\dot{q}_{i}\frac{dt}{d\tau}\right)\frac{d\tau}{dt}\,dt,
\end{equation}
where the proper time interval $d\tau$ is related  to the inertial frame time interval $dt$ by
\begin{equation}
d\tau = dt\sqrt{1 - V^{2}/c^{2}},
\end{equation}
and dots over a quantity denote time derivatives with respect to $t$, e.g., $\dot{q}_{i} = dq_{i}/dt$.  Thus, the Lagrangian for the system relative to the inertial frame is
\begin{equation}
L =  L_{\rm rest}\left(q_{i},\dot{q}_{i}\frac{dt}{d\tau}\right)\frac{d\tau}{dt} = L_{\rm rest}\left[q_{i},q_{i}'(\dot{q}_{i},V)\right] \sqrt{1 - V^{2}/c^{2}}.
\label{L}
\end{equation}
Using Eq.~(\ref{L}), we can obtain the canonical momenta associated with the internal coordinate $q_{i}$,
\begin{eqnarray}
p_{i} & = & \frac{\partial L}{\partial \dot{q}_{i}} \nonumber \\
	& = & \sum_{j}\frac{\partial L_{\rm rest}}{\partial q_{j}'}  \, \frac{dq_{j}'}{d\dot{q}_{i}} \sqrt{1 - V^{2}/c^{2}} \nonumber \\
	& = & \frac{\partial L_{\rm rest}}{\partial q_{i}'}.
\end{eqnarray}
and the center of mass coordinate $X_{i}$, where $X_{1} = X$, $X_{2} = Y$, and $X_{3} = Z$,
\begin{eqnarray}
P_{i} & = & \frac{\partial L}{\partial \dot{X}_{i}} \nonumber \\
	& = &  \sum_{j}\frac{\partial L_{\rm rest}}{\partial q_{j}'} \, \frac{dq_{j}'}{d\dot{X}_{i}} \sqrt{1 - V^{2}/c^{2}}
	  + L_{\rm rest}\frac{\partial}{\partial \dot{X}_{i}}  \sqrt{1 - V^{2}/c^{2}} \nonumber \\
	 	  & = & \left( \sum_{j}\frac{\partial L_{\rm rest}}{\partial q_{j}'} \, q_{j}' - L_{\rm rest}\right) \frac{\dot{X}_{i}/c^{2}}{\sqrt{1 - V^{2}/c^{2}}},
\end{eqnarray}
where we have used $V^{2} = \sum_{k = 1}^{3}\dot{X}_{k}^{2}$.  Since the rest energy of the system is
\begin{equation}
E_{\rm rest} = \sum_{j}\frac{\partial L_{\rm rest}}{\partial q_{j}'} \, q_{j}' - L_{\rm rest},
\end{equation}
we can write the canonical momentum relative to the center of mass velocity $\vec{V}$ as
\begin{equation}
\vec{P} = \frac{(E_{\rm rest}/c^{2})\vec{V}}{\sqrt{1 - V^{2}/c^{2}}}.
\label{P}
\end{equation}
This agrees with the usual expression for a system of mass $M = E_{\rm rest}/c^{2}$.

To obtain the Hamiltonian, we first begin with the constant of motion in terms of coordinates and velocities:
\begin{equation}
h_{0}(X_{i},\dot{X}_{i}; q_{j},\dot{q}_{j}) 	=  \sum_{i = 1}^{3}\dot{X}_{i}P_{i} + \sum_{j}\dot{q}_{j}p_{j} - L = \frac{(E_{\rm rest}/c^{2})}{\sqrt{1 - V^{2}/c^{2}}}.
	\label{h}
\end{equation}
The Hamiltonian is obtained by replacing the coordinate velocities with their associated canonical momenta.  From Eq.~(\ref{P}), we find
\begin{equation}
\frac{1}{\sqrt{1 - V^{2}/c^{2}}} = \frac{\sqrt{E_{\rm rest}^{2}+ P^{2}c^{2}}}{E_{\rm rest}},
\end{equation}
so Eq.~(\ref{h}) becomes
\begin{equation}
H_{0} = \sqrt{E_{\rm rest}^{2} + P^{2}c^{2}},
\label{H02}
\end{equation}
which is the expected answer for a small system with invariant rest energy $E_{\rm rest}/c^{2}$.  Now we write this rest energy as
\begin{equation}
E_{\rm rest} = M_{0}c^{2} + H_{\rm int}(q_{j},p_{j}),
\end{equation}
where all the dynamics of the internal system results from the internal Hamiltonian $H_{\rm int}(q_{j},p_{j})$.  Here we assume that relative to the center of mass, the system is non-relativistic so $H_{\rm int}/M_{0}c^{2} \ll 1$.  Then expanding Eq.~(\ref{H02}) to first order in $H_{\rm int}/M_{0}c^{2}$, we obtain the final result:
\begin{equation}
H_{0}(X_{i},P_{i},q_{j},p_{j}) =  \sqrt{M_{0}^{2}c^{4} + P^{2}c^{2}} + \frac{M_{0}c^{2}H_{\rm int}(q_{j},p_{j})}{\sqrt{M_{0}^{2}c^{4} + P^{2}c^{2}}}.
\label{H03}
\end{equation}

\end{document}